\documentclass[12pt]{article}

\usepackage{apacite,booktabs,multirow,enumitem,multicol,amssymb,amsmath,amsfonts,eurosym,geometry,ulem,graphicx,caption,color,setspace,sectsty,comment,footmisc,caption,natbib,pdflscape,subfigure,array,hyperref}

\newcolumntype{L}[1]{>{\raggedright\let\newline\\arraybackslash\hspace{0pt}}m{#1}}
\newcolumntype{C}[1]{>{\centering\let\newline\\arraybackslash\hspace{0pt}}m{#1}}
\newcolumntype{R}[1]{>{\raggedleft\let\newline\\arraybackslash\hspace{0pt}}m{#1}}

\begin{document}

\begin{titlepage}
\title{Willingness to Pay for an Electricity Connection: A Choice Experiment Among Rural Households and Enterprises in Nigeria\vspace{2em}}
\author{
 Pouya Janghorban\thanks{janghorban@rsm.nl (corresponding address)} \\
  \footnotesize Reiner Lemoine Institute (DE)\vspace{-.5em}\\
  \footnotesize Erasmus University Rotterdam (NL)\\
 \and
 Temilade Sesan\thanks{temi@gbengasesan.com} \\
  \footnotesize University of Ibadan (NG)\\
 \and
 Muhammad-Kabir Salihu\thanks{kabir.salihu@rea.gov.ng} \\
  \footnotesize Rural Electrification Agency (NG)\vspace{-.5em}\\
  \footnotesize Ahmadu Bello University (NG)\\
 \and
 Olayinka Ohunakin\thanks{olayinka.ohunakin@covenantuniversity.edu.ng}\\
  \footnotesize Covenant University (NG)\\
 \and
 Narges Chinichian\thanks{narges.chinichian@rl-institut.de (PI)}\\
 \footnotesize{Reiner Lemoine Institute (DE)}\vspace{-.5em}\\
 \footnotesize Technical University of Berlin (DE)\\
}

\date{\vspace{3em}\today}
\maketitle
\newpage
\begin{abstract}
\noindent Rural electrification initiatives worldwide frequently encounter financial planning challenges due to a lack of reliable market insights. This research delves into the preferences and marginal willingness to pay (mWTP) for upfront electricity connections in rural and peri-urban areas of Nigeria. We investigate discrete choice experiment data gathered from 3,599 households and 1,122 Small to Medium-sized Enterprises (SMEs) across three geopolitical zones of Nigeria, collected during the 2021 PeopleSuN project\footnote{\emph{People Power: Optimizing off-grid electricity supply systems in Nigeria} (PeopleSuN) is a multi-dimensional project involving German and West African researchers and stakeholders. Led by non-profit Reiner Lemoine Institute (Germany) in collaboration with other German partners Technical University Berlin (TUB), Wuppertal Institut, MicroEnergy International (MEI) and Fosera, as well as West African partners: Covenant
University (Nigeria), Obafemi Awolowo University (Nigeria), Université Abdou Moumouni WASCAL program (Niger), Creeds Energy (Nigeria), Clean Technology Hub (Nigeria) and Rural Electrification Agency Nigeria (REA), primarily situated within Nigeria. Spanning a duration of three years, this project has secured funding from the German Federal Ministry of Education and Research. Its primary objective is to address the existing data deficiency pertaining to electricity access in non-urban regions of Nigeria to facilitate the electrification plans by the government, private initiatives and companies, and local community. See here for more information on the project: \href{https://reiner-lemoine-institut.de/en/peoplesun-optimizing-off-grid-electricity-supply-nigeria-2/}{https://reiner-lemoine-institut.de/en/peoplesun-optimizing-off-grid-electricity-supply-nigeria-2/}} survey phase. Employing conditional logit modeling, we analyze this data to explore preferences and marginal willingness to pay for electricity connection. Our findings show that households prioritize nighttime electricity access, while SMEs place a higher value on daytime electricity. When comparing improvements in electricity capacity to medium or high-capacity, SMEs exhibit a sharp increase in willingness to pay for high-capacity, while households value the two options more evenly. Preferences for the electricity source vary among SMEs, but households display a reluctance towards diesel generators and a preference for the grid or solar solutions. Moreover, households with older heads express greater aversion to connection fees, and male-headed households show a stronger preference for nighttime electricity compared to their female-headed counterparts. The outcomes of this study yield pivotal insights to tailor electrification strategies for rural Nigeria, emphasizing the importance of considering the diverse preferences of households and SMEs.\\
\vspace{0in}\\
\noindent\textbf{Keywords:} Willingness to pay, Discrete choice experiment, Rural, Nigeria, Upfront connection fee, Electricity, Off-Grid, Mini-grid Planning\\
\vspace{0in}\\
\noindent\textbf{JEL Codes:} Q41, O13, D12.\\

\end{abstract}
\end{titlepage}

\doublespacing
\setcounter{page}{3 }
\section{Introduction} \label{sec:introduction}
Access to electricity is a crucial aspect of achieving Sustainable Development Goal 7, which aims to "ensure access to affordable, reliable, sustainable, and modern energy for all" \citep{SDG7}. In Nigeria, the most populous country in Africa with over 200 million people, the electrification rate stands at 59.5\%. This leaves about 40\% of the population without proper access to electricity. The disparity is even more pronounced between urban and rural areas, with rural electrification at only 26.3\% in 2021 \citep{world_bank_world_2021}. The inadequacy of electricity access goes beyond just the lack of connection to the national power grid. Even in villages with grid access, residents often face frequent and lengthy power outages \citep{okafor2010challenges, aliyu2013nigeria, pelz_electricity_2023}. As a result, communities turn to alternative backup sources of electricity. The net capacity of all small gasoline generators was reported to be 8 times the grid in Nigeria in 2018 \citep{Generators2019}.

In recent years, various small-scale off-grid and under-grid \citep{Graber2018,Mokoloki} projects have been introduced and implemented in rural Nigeria \citep{NEP}. These initiatives aim to provide a more reliable electricity supply for rural communities, while some also consider enabling them to sell surplus electricity to the national grid in later stages. \cite{herbert2022america} conducted a comparison between the United States' strategy of investing in a dependable nationwide rural grid and India's focus on renewable off-grid solutions. He suggests that the Indian model might offer a more suitable solution for addressing energy poverty in non-urban Nigeria considering its scalability, decentralized approach, and regulatory benefits.

In-depth market insight, financing plans and investment-return strategies in rural areas electrification projects have always been one of the main complications faced by the electrification companies, authorities and rural communities \citep{ohiare_financing_2014, williams2015enabling, o_akpojedje_comprehensive_2016, altawell_rural_2021}. A reliable understanding of the preferences and willingness to pay for electricity among rural communities is needed to conduct accurate feasibility studies and avoid costly mistakes in off-grid planning.\citep{nduka2021get, audu2022expanding}.

We use a choice experiment to evaluate individuals' preferences and willingness to pay for different electricity service attributes. Choice experiment (CE) and contingent valuation (CV) are the two main types of stated choice methods where consumers are surveyed to indirectly assess their preferences and willingness to pay for hypothetical options \citep{breidert_review_2006}. The choice experiment has certain advantages over contingent valuation, such as lower scope insensitivity bias \citep{goldberg2007scope}, and has been used in recent reviews as a benchmark to calibrate different WTP methods results \citep{wang2024household}.

Table \ref{tab:literature} presents a collection of recent studies on willingness to pay in Nigeria, including their covered populations and objectives. While these studies have varied objectives, they all employ contingent valuation methods to estimate willingness to pay, primarily focusing on urban populations or small samples of rural communities \citep{babawale2014estimating, oseni_self-generation_2017, ugulu2019assessing, nduka2023reducing, nduka2021get, onyeuche2021households, ayodele_willingness_2021, audu2022expanding}. 

This study, part of the Project PeopleSuN, contributes to this body of research in three key ways. First, we employ a choice experiment (CE) methodology, which enables a detailed examination of how consumers with different socioeconomic characteristics value various properties of electricity services and their marginal willingness to pay to improve each property. Second, our study is based on a large sample designed to be representative of the rural and peri-urban populations of Nigeria, covering the most extensive areas in these regions and providing higher statistical power. Third, by including small and medium-sized enterprises located outside urban centers, our research goes beyond household surveys and allows for a more comprehensive evaluation of preferences in these enterprises.

This paper is organized as follows: \textbf{Section \ref{sec:method}} details our methodology, including the dataset, the design of the choice experiment, and the econometric model used to analyze the data. Our model evaluates the significance of operational hours (both daytime and nighttime), capacity variations (comparing electricity supply enough for specific appliances versus all devices), the electricity source (national grid, diesel generators, solar microgrids, and solar home systems), and the service provider (local community, private sector or governmental entity). \textbf{Section \ref{sec:result}} presents the results, including models with interaction terms involving socioeconomic factors and gender. The marginal willingness to pay values are also reported in this section. We conclude the paper by discussing the results in \textbf{Section \ref{sec:discussion}} and offering policy recommendations in \textbf{Section \ref{sec:conclusion}}.

\begin{table*}
\centering
\caption{A collection of recent willingness to pay studies in Nigeria}
\scalebox{0.67}{
\begin{tabular}{m{13em} m{2em} m{3em} m{3em} m{14em} m{18em}}
\toprule
Study & Year & Type & Method & Size of Study & Aim \\ 
\midrule

\cite{babawale2014estimating} & 2012$^*$ & Urban &CV & 208 households in 2 residential housing estates in 1 LGA in Lagos & Willingness to pay assessment for improved electricity supply\\

\cite{oseni_self-generation_2017} & 2013 & Urban &CV & 835 households in 2 states in southwest Nigeria: Lagos and Osun, each 3 LGAs & Self-generation influence on willingness to pay for an improved electricity service\\

\cite{onyeuche2021households} & 2018 & Urban &CV & 680 households in total, 170 in each of Abuja, Ibadan, Lagos and Port Harcourt cities & Finding household willingness to pay determinants for improved electricity supply \\

\cite{nduka2021get} & 2018 & Rural &CV & 218  households in 10 LGAs in Ebonyi state &  Willingness to pay estimation for pico-photovoltaic systems and improved cookstoves\\

\cite{nduka2023reducing}& 2018& Urban& CV& 350 households in 10 LGAs in Lagos& Assessing the influence of subsidies on household generator to PV transition\\

\cite{ugulu2019assessing} & 2019$^*$ & Urban &CV & 200 households in all 20 LGAs of Lagos & Willingness to pay assessment for off-grid solar photovoltaic\\

\cite{ayodele_willingness_2021} & 2019 & Urban &CV	& 400 participants from 5 LGAs in Ibadan &  Find factors influencing willingness to Pay for renewable electricity in Nigeria\\

\cite{audu2022expanding} & 2022* & Rural &CV & 400 households in 5 LGAs in Kwara State &  Willingness to pay assessment for Solar Home Systems (SHS)\\


\bottomrule \\ [-1.2ex]

\multicolumn{6}{l}{\emph{Abbreviations:} CV, Contingent Valuation; LGA, Local Government Areas.}\\
\multicolumn{6}{l}{* Not explicitly mentioned in the text.}

\end{tabular}}
\label{tab:literature}

\end{table*}

\section{Methodology} \label{sec:method}
\subsection{Dataset}
\begin{figure}
    \centering
    \includegraphics[width=0.8\linewidth]{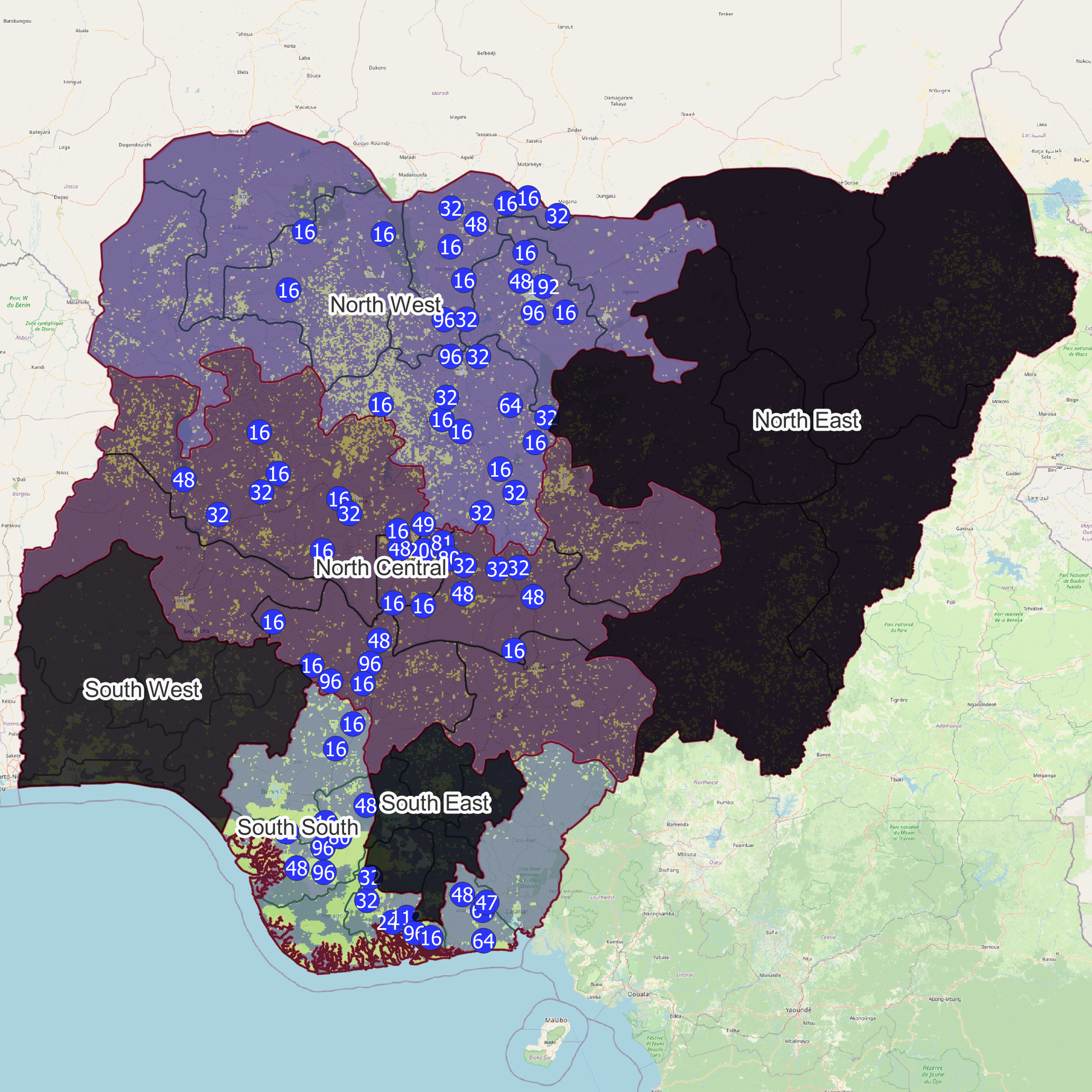}
    \caption{Distribution of sampling locations in three geopolitical zones of Nigeria. Reused from \cite{pelz_electricity_2023}. The background colored map is showing yellow dots as areas with electricity coverage based on 2020 AtlasAI dataset (https://www.atlasai.co, accessed 08.2023). Numbers show how many household surveys were collected in each location. For every 16 households, 5 SMEs were also surveyed. A total number of 3599 households and 1122 SMEs are included in this study. Any inaccuracies in the geographical boundaries are unintentional.}
    \label{fig:Nig_map}
\end{figure}

This research employs answers to several survey questions conducted in a previous stage of the PeopleSuN project. The sample comprises  3599 households and 1122 small and medium-sized enterprises (SMEs) located in grid-electrified rural and peri-urban areas of three different geopolitical zones of Nigeria. These three zones are North West, North Central and South-South. See Figure \ref{fig:Nig_map} for an overview of our surveyed locations. The remaining three geopolitical zones were not included due to high urbanization rate, or safety concerns. See all details regarding the survey and sample design together with the data collection strategies in \cite{pelz_electricity_2023}. Data collection took place in August 2021 and adhered closely to the survey design principles outlined in the World Bank Multi-Tier Framework (MTF) guidelines. Surveys included several modules covering demographic and socioeconomic characteristics, electrical appliances ownership and usage information, electricity access and supply quality assessments, cooking solutions and self-reported capabilities and preferences. Table \ref{tab:summarystats_hh_ent} presents a selected highlight of descriptive statistics from the household and SME samples.
PeopleSuN dataset can be openly accessed on Harvard dataverse\footnote{https://doi.org/10.7910/DVN/GTNEJD}.

\begin{table*}
\centering
\caption{Descriptive summary statistics of the full household and SME samples (N = 3599 and N = 1122 respectively).}
\scalebox{0.67}{
\begin{tabular}[t]{lllrrrrr}
\toprule
Sample & Category & Variable & Mean & SD & Min & Max & N \\
\midrule
\multirow{14}{*}{\rotatebox[origin=c]{90}{Household}}
&\multirow{3}{*}{Demographic}
&Age of household head (years) & 46.81 & 12.51 & 18 & 100 & 3599 \\
&&Gender of household head (m=1,f=2) & 1.11 & 0.31 & 1 & 2 & 3599 \\
&&Household size (no. of people)   & 7.03  & 5.04  & 1  & 57 & 3599 \\ 
\cline{2-8}
&\multirow{5}{*}{Socioeconomic}
&Expenditure (non-energy) weekly (₦) & 12,253.03 & 9,835.02 & 99 & 350,000 & 3595\\
&&Bicycles household owns (count) & 0.27 & 0.65 & 0 & 11 & 3599\\
&&Motorbikes household owns (count) & 0.55 & 0.72 & 0 & 10 & 3599\\
&&Cars household owns (count) & 0.28 & 0.60 & 0 & 8 & 3599\\
&&Bank account  (y=1,n=0) & 0.73 & 0.45 & 0 & 1 & 3599 \\
\cline{2-8}
&\multirow{6}{*}{Electricity connection}
&Any (y=1,n=0) & 1.00 & 0.04 & 0 & 1 & 3599 \\
&&Grid (y=1,n=0) & 0.92 & 0.27 & 0 & 1 & 3599 \\
&&Minigrid (y=1,n=0) & 0.16 & 0.12 & 0 & 1 & 3599 \\
&&Generator (y=1,n=0) & 0.40 & 0.49 & 0 & 1 & 3599 \\
&&Monthly grid bill (if grid main source) (₦) & 3,761 & 3317.61 & 0 & 50,000 & 1968\\
&&Average daily grid blackout (hours) & 18.21 & 5.80 & 0 & 24 & 3316\\

\hline

\multirow{14}{*}{\rotatebox[origin=c]{90}{Enterprise}}
&\multirow{4}{*}{Demographic}
&Years in operation & 7.42 & 5.41 & 2 & 42 & 1122 \\
&&Size (number of full-time employees) & 1.51 & 5.32 & 0 & 99 & 1122 \\
&&Gender of owner (m=1,f=2) & 1.36  & 0.48  & 1  & 2 & 1121 \\
&&Age of owner (year) & 36.96  & 9.37  & 18  & 80 & 1122 
\\ 
\cline{2-8}
&\multirow{4}{*}{Socioeconomic}
&Approximate assets value (₦) & 231,282.90 & 349,630.50 & 2,000 & 5,000,000 & 1118\\
&&Bicycles owned (count) & 0.11 & 0.45 & 0 & 8 & 1122 \\
&&Motorbikes/scooters owned (count) & 0.40 & 0.55 & 0 & 3 & 1122 \\
&&Cars/vans/trucks owned (count) & 0.12 & 0.38 & 0 & 3 & 1122\\
\cline{2-8}
&\multirow{6}{*}{Electricity connection}
&Any (y=1,n=0) & 1.00 & 0.05 & 0 & 1 & 1122 \\
&&Grid (y=1,n=0) & 0.91 & 0.29 & 0 & 1 & 1122 \\
&&Minigrid (y=1,n=0) & 0.02 & 0.14 & 0 & 1 & 1122 \\
&&Generator (y=1,n=0) & 0.58 & 0.49 & 0 & 1 & 1122 \\
&&Monthly grid bill (if grid main source) (₦) & 3,036.55	 & 3,634.32 & 99 & 70,000 & 1011\\
&&Average daily grid blackout (hours) & 18.44 & 5.04 & 0 & 24 & 1018\\
\bottomrule
\end{tabular}}
\label{tab:summarystats_hh_ent}
\end{table*}

\subsection{Design of choice experiment}
\begin{figure*}
    \centering
    \includegraphics[width=.8\textwidth]{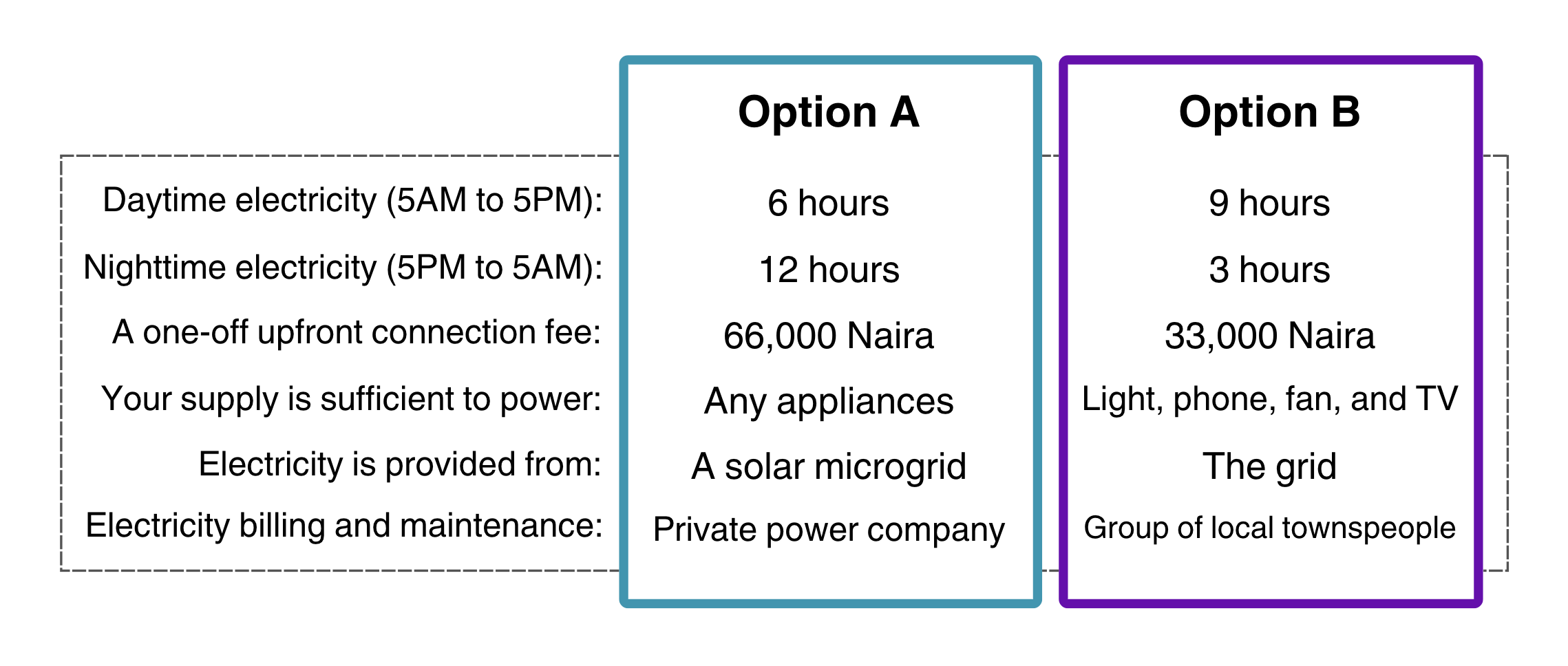}
    \caption{A choice task example with two choice cards and no opt-out option. The choice tasks have been translated into the local languages for user accessibility.}
    \label{fig:choice_card_example}
\end{figure*}

Survey participants were requested to envision a scenario where they were relocating to a new building without electricity and needed to choose a new service. They were then presented with choice cards, similar to the example depicted in Figure \ref{fig:choice_card_example}, requiring them to decide between options A and B. These options were configured with attributes randomized through an orthogonal fractional factorial design. Each survey respondent completed two choice tasks. Two households and one enterprise had no task completed and 104 households and 26 enterprises had only one task performed. We excluded the households and enterprises with no answer and kept the ones with single responses (a total of 7090 tasks for households and 2216 tasks for SMEs). Enumerators clarified to respondents that they needed to select either option A or B based on their preferences for the displayed attributes, with the two options being independently randomized and distinct each time.

A comprehensive list of attributes, their descriptions, and respective levels can be found in Table \ref{tab:choice_attributes}. The selection of these attributes was for the most part influenced by the multi-tier framework (MTF) introduced by \cite{world_bank_beyond_2015}. Collaboration with PeopleSuN project partners in Nigeria further informed the choice of attributes, catering to their specific inquiries. To ensure survey consistency and alleviate respondent cognitive burden, the inclusion of additional attributes within this discrete choice experiment was constrained \citep{carson1994experimental}. Therefore, after consultation with study partners, we prioritized and selected the most crucial attributes. This approach aimed to balance the need to collect further information with the goal of providing a survey experience that was easy to understand and manage for respondents.

\begin{table}
\centering
\caption{List of the choice experiment attributes implemented in this study with respective descriptions and levels.}
\scalebox{0.75}{
\begin{tabular}{m{8em} m{4em} m{18em} m{18em}}
\toprule
Attribute & Sign & Description & Levels \\ \midrule
Daytime hours & am & On average, how many hours of electricity in the daytime (5AM to 5PM) is guaranteed & 
\begin{itemize}[leftmargin=*, noitemsep]
    \item 3 hours
    \item 6 hours
    \item 9 hours
    \item 12 hours
\end{itemize}\\ 
Night-time hours   & pm   & On average, how many hours of electricity in the nightime (5PM to 5AM) is guaranteed  & 
\begin{itemize}[leftmargin=*, noitemsep]
    \item 3 hours
    \item 6 hours
    \item 9 hours
    \item 12 hours
\end{itemize}\\ 
Fee           & fee   & The upfront connection fee (including in-home wiring   and connection) & 
\begin{itemize}[leftmargin=*, noitemsep]
    \item ₦3,000
    \item ₦33,000
    \item ₦66,000
    \item ₦100,000
\end{itemize}\\ 
Capacity     & cap  & The supply capacity is sufficient to power which appliances & 
\begin{itemize}[leftmargin=*, noitemsep]
    \item low: Lighting and phone charging
    \item mid: Lighting, phone charging, fan, TV
    \item high: Any appliances
\end{itemize}\\ 
Source          & src   & The generation source of the electricity & 
\begin{itemize}[leftmargin=*, noitemsep]
    \item grid: The national grid
    \item smg: A solar micro-grid
    \item shs: A solar home system
    \item diesel: A diesel generator
\end{itemize}\\ 
Carrier         & crr   & Electricity billing and maintenance management   organization & \begin{itemize}[leftmargin=*, noitemsep]
    \item prv: A private company
    \item stt: The state electricity company
    \item com: A group of local townspeople
\end{itemize}\\ \bottomrule
\end{tabular}}
\label{tab:choice_attributes}
\end{table}

\subsection{Econometric model}

The random utility theory \citep{mcfadden19741973,louviere_stated_2000} provides a framework to model choice experiment data and gain insight into individuals' preferences and willingness to pay. This framework operates on the foundational assumptions of economic rationality and utility maximization. It is postulated that an individual's preferences among available alternatives can be captured through a utility function and the individual chooses the option with the highest utility after evaluating all available options. The total utility that an individual can derive from an alternative $i$ in a choice set is denoted as $U_i$, which can be broken down into a linear composition of those contributions observed by the researcher ($V_i$) as well as the unobserved contributions ($\epsilon_i$). Thus, $U_i=V_i+\epsilon_i$. It is common to express $V_i$ as a linear composition of the attributes in the choice experiment, i.e. $V_i=\sum_{j=1}^K \beta_j X_j$, where $X_j$ denote each attribute and $\beta_j$ are the respective coefficients. Categorical attributes need to be coded into levels using methods such as dummy coding to enter the utility function \citep{mariel_environmental_2021}. 

In this study, the three attributes of upfront connection fee (fee), daytime hours (am), and nighttime hours (pm) are continuous variables and enter the utility function directly. The other three attributes are categorical variables. The capacity attribute (cap) is dummy coded into cap\_mid and cap\_high, with the low capacity as the reference level. Further, three dummy variables, namely src\_shs (solar home system), src\_smg (solar micro-grid), src\_grid (the grid) are introduced in the process of coding the categorical attribute source (src) with the diesel generator as the reference level. Likewise, the variable carrier (crr) is coded into crr\_prv (private carrier) and crr\_com (community-run carrier by a group of local townspeople) with the stated-owned carrier as the reference level. The resulting utility function for this study is:
\begin{multline} \label{eqn:full_utility}
    U_i = \beta_{1}\cdot\textrm{fee} + \beta_{2}\cdot\textrm{am} + \beta_{3}\cdot\textrm{pm} + \beta_{4}\cdot\textrm{cap\_mid} \\+ \beta_{5}\cdot\textrm{cap\_high} + \beta_{6}\cdot\textrm{src\_shs} + \beta_{7}\cdot\textrm{src\_smg} + \beta_{8}\cdot\textrm{src\_grid} \\+ \beta_{9}\cdot\textrm{crr\_prv} + \beta_{10}\cdot\textrm{crr\_com} + \epsilon_i   
\end{multline}

 The unobserved components of the utility ($\epsilon_i$) are treated as stochastic influences. In Conditional Logit Modeling, it is assumed that $\epsilon_i$ are \emph{independent and identically distributed} (IID) across individuals, alternatives, and choice situations, with an \emph{extreme value type 1} (EV1, a.k.a. \emph{Gumbel}) distribution \citep{hensher_applied_2015}. Given these assumptions, the probability of choosing alternative $i$ among all the other alternatives can be derived as \citep[p. 42-47]{louviere_stated_2000}:
\begin{equation} \label{eqn:mnl}
    P_i = \frac{e^{V_i}}{\sum_{j=1}^{J}e^{V_j}} \qquad (j=1,2,\cdots,J) \quad (i\neq j)
\end{equation}
where $\sum_{j=1}^{J}e^{V_j}$ denotes the sum of the observed components of the utilities of all alternatives ($V_j$s) except the alternative that we are calculating its probability ($V_i$). Given this relation and the choice experiment data, the coefficients of the utility function ($\beta_i$) can be estimated using mathematical analysis methods such as the \emph{maximum likelihood method} \citep[p. 47]{louviere_stated_2000}. In this study, we used the 'mlogit' package in R \citep{croissant_estimation_2020} to calculate the coefficients of our model.

Marginal willingness to pay values can be calculated using the coefficients of the utility function:  
\begin{equation} \label{eqn:wtp_formula}
    \textrm{mWTP}_\textrm{attribute}=-\frac{\beta_\textrm{attribute}}{\beta_\textrm{price}}
\end{equation}
This value indicates how much individuals are willing to pay on average to get a service with one unit improvement in the respective attribute (for continuous variables) or to get that specific level instead of the reference level (in the case of dummy variables).

Moreover, the model can incorporate socioeconomic factors by introducing interaction terms with the service attributes. Analyzing the coefficients of these interaction terms can provide insights into how specific socioeconomic factors influence individuals' preferences for a service attribute. In this study, we added interaction terms between the electricity connection fee (variable fee) and eight socioeconomic factors of the household: weekly expenditure per member (weekly expenditure of household divided by household size), age of the head, marital status of the head (married/single, separated, divorced), gender of the head, a measure of gender dynamics at the household level (answer to the survey question: \emph{Do the lead women in the household need to ask others for permission before buying clothing for themselves? [Likert] Never/Rarely/Yes,if unmarried/Yes,usually/Yes,always}), household size, previous loan experience (answer to survey question: \emph{Has anyone in your family taken a loan to buy a household appliance? Yes/No}), and Trust in government (answer to the survey question: \emph{How much trust do you have in government authorities in general? [Likert] Not at all/Not very much/Somewhat/A great deal}).

\section{Results} \label{sec:result}
\subsection{Household and enterprise preferences}
The results of the conditional logit modeling for the households and enterprises are presented in Table \ref{tab:hh_mnl_results} and Table \ref{tab:ent_mnl_results} respectively. Four different models have been estimated and reported for each sample in the columns. 

\begin{table}
\centering
\caption{Household sample coefficient estimates of the conditional logit model. The estimated means of the coefficients are reported in each row. The standard deviation of each coefficient is noted in parenthesis below its estimated mean. Models {\footnotesize(1)}, {\footnotesize(2)}, and {\footnotesize(3)} progressively include further attributes of the electricity service into consideration, with model {\footnotesize(3)} following the full utility function presented in equation \ref{eqn:full_utility}. Model {\footnotesize(4)} only differs from model {\footnotesize(3)} in the choice of the dummy variable src (source of electricity) base level: Model {\footnotesize(3)} base is set to the national grid, whereas in model {\footnotesize(4)}, the diesel generator is assumed to be the base.}
\scalebox{0.8}{
\begin{tabular}{@{\extracolsep{2pt}}lcccc} 
\\[-1.8ex]\hline 
\hline \\[-1.8ex] 
 & \multicolumn{4}{c}{\textit{Models}} \\ 
\cline{2-5} 
\\[-1.8ex] Coefficient & (1) & (2) & (3) & (4)\\ [.8ex] 
\hline \\[-1ex] 
 fee & $-$0.042$^{***}$ & $-$0.042$^{***}$ & $-$0.042$^{***}$ & $-$0.042$^{***}$ \\ 
  & (0.001) & (0.001) & (0.001) & (0.001) \\ 
  & & & & \\ 
 am & 0.072$^{***}$ & 0.072$^{***}$ & 0.072$^{***}$ & 0.072$^{***}$ \\ 
  & (0.007) & (0.007) & (0.007) & (0.007) \\ 
  & & & & \\ 
 pm & 0.085$^{***}$ & 0.086$^{***}$ & 0.086$^{***}$ & 0.086$^{***}$ \\ 
  & (0.007) & (0.007) & (0.007) & (0.007) \\ 
  & & & & \\ 
 cap\_mid  & 0.651$^{***}$ & 0.654$^{***}$ & 0.653$^{***}$ & 0.653$^{***}$ \\ 
    & (0.058) & (0.058) & (0.058) & (0.058) \\ [-2ex] 
  & & & & \\ 
 cap\_high  & 1.336$^{***}$ & 1.338$^{***}$ & 1.336$^{***}$ & 1.336$^{***}$ \\ 
    & (0.062) & (0.062) & (0.062) & (0.062) \\ 
  & & & & \\ 
 src\_shs  &  & 0.193$^{***}$ & 0.035 & 0.192$^{***}$ \\ 
    &  & (0.065) & (0.065) & (0.065)\\ [-2ex]
  & & & & \\ 
 src\_smg &   & 0.195$^{***}$ & 0.041 & 0.197$^{***}$ \\  
    &  & (0.066) & (0.066) & (0.066) \\ [-2ex]
  & & & & \\ 
 src\_grid  &  & 0.153$^{**}$ &  & 0.156$^{**}$ \\ 
    &  & (0.065) & & (0.065) \\ [-2ex]
  & & & & \\ 
 src\_diesel &  &  & $-$0.156$^{**}$ &  \\ 
  &  &  & (0.065) & \\ 
  & & & & \\ 
 crr\_prv &  &  & 0.098 & 0.098 \\ 
  &  &  & (0.057) & (0.057) \\ [-2ex]
  & & & & \\ 
 crr\_com &  &  & 0.083 & 0.083 \\ 
  &  &  & (0.056) & (0.056) \\ 
  & & & & \\ 
\hline \\[-1.8ex] 
Observations & 7,090 & 7,090 & 7,090 & 7,090 \\ 
Log Likelihood & $-$2,906.090 & $-$2,900.166 & $-$2,898.433 & $-$2,898.433 \\ 
$\rho^2$ & 0.4086 & 0.4098 & 0.410141 & 0.410141\\ 
AIC & 5822.18 & 5816.33 & 5816.87 & 5816.87 \\ 
\hline
\hline \\[-1.8ex] 
\textit{Note:}  & \multicolumn{4}{r}{$^{*}$p$<$0.1; $^{**}$p$<$0.05; $^{***}$p$<$0.01} \\ 
\end{tabular} }
\label{tab:hh_mnl_results}
\end{table}

\begin{table}
\centering
\caption{Enterprise sample coefficient estimates of the conditional logit model. The estimated means of the coefficients are reported in each row. The standard deviation of each coefficient is noted in parenthesis below its estimated mean. Models {\footnotesize(1)}, {\footnotesize(2)}, {\footnotesize(3)}, and {\footnotesize(4)} progressively include further attributes of the electricity service into consideration, with model {\footnotesize(4)} following the full utility function presented in equation \ref{eqn:full_utility}.}
\scalebox{0.8}{
\begin{tabular}{@{\extracolsep{5pt}}lcccc} 
\\[-1.8ex]\hline 
\hline \\[-1.8ex] 
 & \multicolumn{4}{c}{\textit{Models}} \\ 
\cline{2-5} 
\\[-1.8ex] Coefficient & (1) & (2) & (3) & (4)\\ [.8ex]
\hline \\[-1ex] 
 fee & $-$0.033$^{***}$ & $-$0.036$^{***}$ & $-$0.036$^{***}$ & $-$0.036$^{***}$ \\ 
  & (0.001) & (0.002) & (0.002) & (0.002) \\ 
  & & & & \\ 
 am & 0.063$^{***}$ & 0.076$^{***}$ & 0.076$^{***}$ & 0.076$^{***}$ \\ 
  & (0.011) & (0.012) & (0.012) & (0.012) \\ 
  & & & & \\ 
 pm & 0.033$^{***}$ & 0.039$^{***}$ & 0.039$^{***}$ & 0.039$^{***}$ \\ 
  & (0.011) & (0.012) & (0.012) & (0.012) \\ 
  & & & & \\ 
 cap\_mid &  & 0.461$^{***}$ & 0.466$^{***}$ & 0.464$^{***}$ \\ 
  &  & (0.097) & (0.097) & (0.098) \\ [-2ex]
  & & & & \\ 
 cap\_high &  & 1.232$^{***}$ & 1.235$^{***}$ & 1.233$^{***}$ \\ 
  &  & (0.103) & (0.103) & (0.103) \\ 
  & & & & \\ 
 src\_shs &  &  & 0.109 & 0.110 \\ 
  &  &  & (0.112) & (0.112) \\ [-2ex]
  & & & & \\ 
 src\_smg &  &  & 0.136 & 0.137 \\ 
  &  &  & (0.113) & (0.114) \\ [-2ex]
  & & & & \\ 
 src\_grid &  &  & 0.118 & 0.121 \\ 
  &  &  & (0.111) & (0.111) \\ 
  & & & & \\ 
 crr\_prv &  &  &  & $-$0.097 \\ 
  &  &  &  & (0.094) \\ [-2ex]
  & & & & \\ 
 crr\_com &  &  &  & $-$0.122 \\ 
  &  &  &  & (0.097) \\ 
  & & & & \\ 
\hline \\[-1.8ex] 
Observations & 2,216 & 2,216 & 2,216 & 2,216 \\ 
Log Likelihood & $-$1,077.756 & $-$996.152 & $-$995.246 & $-$994.356 \\ 
$\rho^2$ & 0.2972 & 0.3504 & 0.3510 &  0.3516 \\ 
AIC & 2161.51 & 2002.30 & 2006.49 & 2008.71 \\ 
\hline 
\hline \\[-1.8ex] 
\textit{Note:}  & \multicolumn{4}{r}{$^{*}$p$<$0.1; $^{**}$p$<$0.05; $^{***}$p$<$0.01} \\ 
\end{tabular}}
\label{tab:ent_mnl_results}
\end{table}

For the households, most attributes have high significance at 0.01 and 0.05, except for the choice of electricity carrier (crr\_prv and crr\_com coefficients), as well as the source of electricity when the grid is taken as the reference level (refer to model (3) in Table \ref{tab:hh_mnl_results} as opposed to model (4) in this table). For the enterprises, however, the coefficients of electricity carrier (crr\_prv and crr\_com) and source of electricity (src\_shs, src\_smg, and src\_grid) both are not significantly different from zero. Except for the fee coefficient in both samples, all the other statistically significant coefficients have positive estimations. Positive coefficients suggest an increase in utility and preference, while negative values suggest the opposite. Therefore, our estimated coefficients indicate that households and enterprises prefer a service with a lower connection fee, with more operational hours (am and pm), and with medium or high-capacity electricity (cap\_mid and cap\_high) instead of low-capacity. Moreover, households prefer electricity sourced from a solar solution or the grid (src\_shs, src\_smg, and src\_grid) as an improvement to the diesel generator solution.

\subsection{Willingness to pay}

Given the coefficients estimated in the Table \ref{tab:hh_mnl_results} and Table \ref{tab:ent_mnl_results}, we used the formula in equation \ref{eqn:wtp_formula} to calculate the marginal willingness to pay (mWTP) values for the upfront electricity connection fee. Coefficients from model (3) for households and model (2) for enterprises were used to calculate mWTP values due to their lower Akaike Information Criterion (AIC) value (better goodness of fit). The results are presented in Table \ref{tab:wtp}. It can be seen that the mWTP value for daytime electricity hours is higher among enterprises compared to households, with respectively ₦2,115 and ₦1,709. However, this trend reverses for nighttime electricity, with households willing to pay ₦2,025 for each extra hour of nighttime guaranteed electricity compared to ₦1,076 for the enterprises. 

The attribute of electricity capacity yielded the highest willingness to pay amounts, with households and enterprises demonstrating willingness to pay upfront ₦15,458 and ₦12,833, respectively, for transitioning to an electricity service with medium capacity (capable of powering lighting, phone charging, fan, and TV) instead of low capacity (suitable only for lighting and phone charging). These figures would rise to ₦31,603 for households and ₦34,282 for enterprises when transitioning from low to high-capacity electricity capable of operating any household appliance. Once again, a disparity between households and enterprises is evident. While households are more inclined to pay for a medium-capacity service than enterprises (₦15,458 compared to ₦12,833), high-capacity electricity holds greater appeal for enterprises, reflecting a higher marginal willingness to pay, amounting to ₦34,282 compared to ₦31,603 of the households.

Households showed a higher preference for solar solutions over grid-based electricity as an alternative to diesel generators. The mWTP for adopting a solar microgrid or solar home system instead of a diesel generator was nearly the same (₦4,613 and ₦4,550), exceeding the mWTP for transitioning to grid electricity from a diesel generator by almost a thousand Naira. Thus, we can rank electricity solutions in terms of household appeal over diesel generators: solar microgrid (highest appeal), solar home system, and the national grid (least appeal). Determining an overall enterprise mWTP values for this attribute wasn't possible, given that the corresponding coefficients in Table \ref{tab:ent_mnl_results} were not significantly different from zero.

\begin{table}
\caption{Estimated marginal willingness to pay values.}
\centering
\scalebox{0.8}{
\begin{tabular}{m{4em} p{24em} p{6em} p{6em}}
\\[-1.8ex]
\hline \hline \\ [-1.6ex]
    Variable & Improvement & Households & Enterprises \\ [.5ex]
    \hline\\[-1.8ex]
    am & Each extra hour of daytime guaranteed electricity & ₦1,709  & ₦2,115 \\ 
    pm & Each extra hour of night-time guaranteed electricity & ₦2,025 & ₦1,076 \\ 
    cap\_mid & Electricity capacity from low to mid & ₦15,458 & ₦12,833 \\
    cap\_high & Electricity capacity from low to high & ₦31,603  & ₦34,282 \\
    src\_shs & Electricity source from Diesel to Solar Home System & ₦4,550 & -\\ 
    src\_smg & Electricity source from Diesel to Solar Micro-Grid & ₦4,613 & -\\ 
    src\_grid & Electricity source from Diesel to the [national] Grid & ₦3,616 & -\\[1ex]
\hline \hline \\ [-1ex]
    \multicolumn{4}{l}{\emph{Note:} Values are in ₦ (Nigerian Naira). In 2021, the exchange rate was approximately ₦400=\$1.}\\
\end{tabular}}
\label{tab:wtp}
\end{table}

\subsection{Interaction terms with socioeconomic factors of the households}

\begin{table}
\caption{Household conditional logit model including interaction terms between socioeconomic factors and electricity connection fee.}
\centering
\scalebox{0.78}{
\begin{tabular}{m{.1cm}lp{2cm}p{1cm}} 
\hline 
\hline \\[-4.5ex] 
\\& & Mean & $\sigma$ \\ [.5ex]
\hline \\[-1.8ex] 
\multicolumn{2}{l}{Electricity attributes} &\\[1.8ex]
& fee & $-$0.040$^{***}$ & 0.006 \\ [1ex]
& am & 0.075$^{***}$ & 0.007 \\ [1ex]
& pm & 0.088$^{***}$ & 0.007 \\ [1ex]
& cap\_mid & 0.693$^{***}$ & 0.060 \\ [1ex]
& cap\_high & 1.391$^{***}$ & 0.064 \\ [1ex]
& src\_shs & 0.221$^{***}$ & 0.067 \\ [1ex]
& src\_smg & 0.234$^{***}$ & 0.068 \\ [1ex]
& src\_grid & 0.180$^{**}$ & 0.067 \\ [1ex]
& crr\_prv & 0.093 & 0.059 \\ [1ex]
& crr\_com & 0.078 & 0.058 \\ [1.8ex]
\multicolumn{2}{l}{Interaction terms with fee} &\\ [1.5ex]
& Weekly expenditure per member & 0.0019$^{***}$ & 0.00037 \\ [1ex]
& Age of household head & $-$0.0003$^{***}$ & 0.00009 \\ [1ex]
& Married household head & 0.0053 & 0.00339 \\ [1ex]
& Male household head & $-$0.0030 & 0.00332 \\ [1ex]
& Female need permission buying clothes & $-$0.0019$^{**}$ & 0.00067 \\ [1ex]
& Household size & $-$0.0001 & 0.00025 \\ [1ex]
& Taken loan(s) before & $-$0.0031 & 0.00948 \\ [1ex]
& Trust in government & 0.0038$^{***}$ & 0.00101 \\ [1ex]
\hline \\[-1.2ex] 
\multicolumn{2}{l}{Goodness of fit} && \\ [1ex]
& Number of observations & \multicolumn{2}{c}{6,917} \\ 
& Log Likelihood & \multicolumn{2}{c}{$-$2,760.619} \\ 
& $\rho^2$ & \multicolumn{2}{c}{0.4242} \\ 
& AIC & \multicolumn{2}{c}{5557.237} \\
\hline 
\hline \\[-1.8ex] 
\textit{Note:}  && \multicolumn{2}{r}{$^{*}$p$<$0.1; $^{**}$p$<$0.05; $^{***}$p$<$0.01} \\ 
\end{tabular}}
\label{tab:interaction}
\end{table}

The results of the household model with interaction terms are given in Table \ref{tab:interaction}. A number of socioeconomic characteristics are selected from the full survey data to be studied in relation to the fee.  The number of observations in this model is less than the previous models presented due to a number of missing answers to the socioeconomic questions of the survey. The coefficient of weekly expenditure per household member is significantly different from zero and positive. This indicates that households with higher expenditure per member have less aversion to paying higher electricity connection fees.

Only the coefficient of age was found to be statistically significant among the household head characteristics we assessed (i.e. age, marital status, and gender). Its negative sign conveys that households with older heads are more averse to paying higher connection fees. 
While household size and previous experience of taking loans for buying appliances were not statistically significant in our model, we found the coefficient for trust in the government to be significantly different from zero and positive, conveying that households with higher trust in the government are willing to pay higher amounts for the electricity connection fee. What is more, we found the term indicating the extent to which the lead women in the household need permission before buying clothes for themselves to have a statistically significant and negative interaction with the variable fee. This suggests that these households are more averse to paying higher electricity connection fees. These findings indicate that family dynamics and gender aspects of willingness to pay 
shape an important dimension of our overall understanding in this topic. The following subsection focuses on gender-related aspects of our data.

\subsubsection{Gender-related interaction terms}

\begin{table*}
\caption{Household conditional logit model including gender-related interaction terms.}
\centering
\scalebox{0.7}{
\begin{tabular}{@{\extracolsep{5pt}}lcccc} 
\\[-1.8ex]\hline 
\hline \\[-1.8ex] 
 & \multicolumn{4}{c}{\textit{Regions}} \\
\cline{2-5}\\ [-1ex]
 & North Central & South South & North West & All \\ 
\hline \\[-1.8ex] 
 fee & $-$0.077$^{***}$ & $-$0.031$^{***}$ & $-$0.056$^{***}$ & $-$0.036$^{***}$ \\ 
  & (0.011) & (0.004) & (0.012) & (0.003) \\ 
  & & & & \\ 
 am & 0.211$^{***}$ & 0.023 & 0.013 & 0.058$^{***}$ \\ 
  & (0.059) & (0.029) & (0.076) & (0.022) \\ 
  & & & & \\ 
 pm & 0.062 & 0.027 & 0.128$^{*}$ & 0.046$^{**}$ \\ 
  & (0.051) & (0.028) & (0.075) & (0.021) \\ 
  & & & & \\ 
 cap\_mid & 0.902$^{*}$ & 1.170$^{***}$ & 0.159 & 0.845$^{***}$ \\ 
  & (0.471) & (0.252) & (0.563) & (0.186) \\ [-2ex]
  & & & & \\ 
 cap\_high & 1.230$^{***}$ & 2.146$^{***}$ & 0.627 & 1.474$^{***}$ \\ 
  & (0.460) & (0.274) & (0.604) & (0.194) \\ 
  & & & & \\ 
 src\_shs & 0.327$^{***}$ & 0.087 & 0.284$^{**}$ & 0.200$^{***}$ \\ 
  & (0.121) & (0.110) & (0.117) & (0.066) \\ [-2ex]
  & & & & \\ 
 src\_smg & 0.107 & 0.184$^{*}$ & 0.323$^{***}$ & 0.196$^{***}$ \\ 
  & (0.125) & (0.111) & (0.116) & (0.066) \\ [-2ex]
  & & & & \\ 
 src\_grid & 0.149 & 0.157 & 0.174 & 0.153$^{**}$ \\ 
  & (0.123) & (0.111) & (0.115) & (0.065) \\
  & & & & \\ 
 crr\_prv & 0.093 & 0.074 & 0.168$^{*}$ & 0.100$^{*}$ \\ 
  & (0.107) & (0.096) & (0.102) & (0.057) \\ [-2ex]
  & & & & \\ 
 crr\_com & 0.202$^{*}$ & $-$0.024 & 0.128 & 0.089 \\ 
  & (0.107) & (0.095) & (0.099) & (0.057) \\ 
  & & & & \\ 
 am:gender\_head\_male & $-$0.109$^{*}$ & 0.048$^{*}$ & 0.055 & 0.026 \\ 
  & (0.059) & (0.028) & (0.074) & (0.022) \\ [-2ex]
  & & & & \\ 
 pm:gender\_head\_male & 0.039 & 0.063$^{**}$ & $-$0.003 & 0.051$^{**}$ \\ 
  & (0.052) & (0.028) & (0.075) & (0.022) \\ [-2ex]
  & & & & \\ 
 fee:gender\_head\_male & 0.025$^{**}$ & $-$0.011$^{***}$ & 0.021$^{*}$ & $-$0.004 \\ 
  & (0.011) & (0.004) & (0.012) & (0.003) \\ [-2ex]
  & & & & \\ 
 cap\_mid:gender\_head\_male & $-$0.637 & 0.236 & 0.088 & $-$0.079 \\ 
  & (0.470) & (0.245) & (0.548) & (0.188) \\ [-2ex]
  & & & & \\ 
 cap\_high:gender\_head\_male & $-$0.339 & 0.734$^{***}$ & 0.380 & 0.269 \\ 
  & (0.461) & (0.276) & (0.596) & (0.199) \\ 
  & & & & \\ 
 am:female\_permission\_clothes & $-$0.001 & 0.004 & $-$0.002 & $-$0.003 \\ 
  & (0.008) & (0.010) & (0.008) & (0.005) \\ [-2ex]
  & & & & \\ 
 pm:female\_permission\_clothes & 0.009 & $-$0.004 & $-$0.013 & $-$0.002 \\ 
  & (0.009) & (0.010) & (0.008) & (0.005) \\ [-2ex]
  & & & & \\ 
 fee:female\_permission\_clothes & 0.001 & 0.002 & $-$0.003$^{**}$ & $-$0.001$^{*}$ \\ 
  & (0.001) & (0.001) & (0.001) & (0.001) \\ [-2ex]
  & & & & \\ 
 cap\_mid:female\_permission\_clothes & $-$0.006 & $-$0.158$^{*}$ & 0.112 & $-$0.042 \\ 
  & (0.068) & (0.083) & (0.068) & (0.037) \\ [-2ex]
  & & & & \\ 
 cap\_high:female\_permission\_clothes & $-$0.062 & $-$0.397$^{***}$ & 0.099 & $-$0.135$^{***}$ \\ 
  & (0.071) & (0.090) & (0.075) & (0.040) \\ 
  & & & & \\ 
\hline \\[-1.8ex] 
Observations & 2,368 & 2,359 & 2,363 & 7,090 \\ 
Log Likelihood & $-$860.409 & $-$996.430 & $-$926.352 & $-$2,882.620 \\ 
$\rho^2$ & 0.4757 & 0.3906 & 0.4344 & 0.4134 \\
AIC & 1760.82 & 2032.86 & 1892.70 & 5805.24 \\
\hline 
\hline \\[-1.8ex] 
\textit{Note:}  & \multicolumn{4}{r}{$^{*}$p$<$0.1; $^{**}$p$<$0.05; $^{***}$p$<$0.01} \\ 
\end{tabular} }
\label{tab:wtp_gender_interaction}
\end{table*}

Table \ref{tab:wtp_gender_interaction} presents the influence of household heads' gender on their preference for each of the electricity service attributes and overall willingness to pay. The count of male/female-headed households in our surveys is presented in Figure \ref{fig:male_female_percentage}. On the national level (all three zones aggregated) we found only the influence of the household head gender on the preference for nighttime electricity to be statistically significant: Male-headed households showed a higher preference for increased nighttime electricity.
However, preference for other attributes, including capacity and willingness to pay were not significantly influenced by the gender of the household head.

\begin{table*}
  \centering
  \caption{Female-headed households vs male-headed households expenditure in ₦ by zone (SD= Standard Deviation)}
  \label{tab:expenditure}
\scalebox{0.8}{
  \begin{tabular}{lccc}
    \toprule
    Zone & Average Female Expenditure & Average Male Expenditure & Ratio Female/Male \\
    \midrule
    North Central & 8,946 {\footnotesize(SD: 5,836)} & 10,444 {\footnotesize(SD:8,060)} & 0.86 \\
    South South & 10,644 {\footnotesize(SD:5,759)} & 14,526 {\footnotesize(SD:7,340)} & 0.73 \\
    North West & 8,603 {\footnotesize(SD:6,069)} & 12,853 {\footnotesize(SD:13,254)} & 0.67 \\
    \bottomrule
  \end{tabular}}
\end{table*}

Keeping the heterogeneity of the Nigerian population in mind, at the zonal level, we observe different patterns in each zone. While in North West we could not detect any influence of household head gender on their preferences for electricity service, in South South and North Central, this factor had a statistically significant influence on the household WTP. In South South, male-headed households show a higher aversion to expensive electricity connection compared to female-headed households as well as a higher preference for increased nighttime electricity hours. On the other hand, in North Central, male-headed households show less aversion to expensive electricity connection fee compared to female-headed households. The influence of gender on preference for other factors was not found statistically significant.

\section{Discussion} \label{sec:discussion}
\subsection{Pronounced high-capacity daytime electricity demand among SMEs}
The assessment of willingness to pay values between daytime and nighttime electricity hours reveals a distinct pattern: households prioritize nighttime electricity, whereas enterprises place a higher value on daytime electricity. However, the divergence in marginal WTP levels for daytime and nighttime electricity is considerably more pronounced among enterprises, exceeding threefold when compared to only doubled values for households. This observation may indicate that household electricity demand is more evenly spread across the entire 24-hour cycle, whereas SMEs exhibit a substantial demand concentration during daytime hours.

Our study revealed a significant difference between households and SMEs in terms of their marginal WTP for improving the capacity attribute. Specifically, we observed a substantial disparity in the mWTP to improve from low to medium-capacity electricity, as compared to an improvement from low to high-capacity electricity. This was particularly pronounced among SMEs in comparison to households. The discrepancy suggests that compared to SMEs, households would generally find medium-capacity electricity more satisfactory. SMEs, however, appear to place significant value on high-capacity electricity capable of powering all appliances. This inclination aligns with the typical electricity demands of businesses, extending beyond basic functions like lighting, phone charging, fan, and TV to encompass more extensive requirements such as refrigeration and operation of electrical machinery.

\subsection{Households' aversion to diesel generators}
We observed that households prefer to switch from diesel generators to alternative electricity sources. This aversion to generators, reported in previous studies in Nigeria, may stem from various reasons. \cite{oseni_self-generation_2017} found that households with generators are more willing to pay for reliable grid electricity due to the high costs of self-generation. Additionally, generators cause air and noise pollution and present challenges in procuring and transporting fuel, especially in rural areas \citep{world_bank_dirty_2019}.

\cite{nduka2023reducing} noted a higher tendency for generator-owning households to switch to solar solutions, though this may not be driven by environmental concerns. Our assessment of preferences for switching from the grid to solar electricity (see model (3) in Table 2) did not yield statistically significant results. The trade-off between the challenges of the grid supply (e.g., low reliability and scarcity in rural areas) and those of the solar solutions (e.g., lower capacity, maintenance challenges; see \cite{berger2017practical} for a detailed account) does not have a clear winner the households' perspective.

Among SMEs, we did not observe any statistically significant preference for a particular source of electricity generation. The electricity needs of SMEs vary depending on different conditions and requirements. Consequently, adopting a one-size-fits-all approach to SMEs’ electricity preferences is not feasible. Instead, SMEs may opt for a combination of energy sources to ensure uninterrupted operations, taking into account their unique requirements. Moreover, SMEs are more inclined than households to conduct a cost-benefit analysis of their electricity options.

However, it is crucial to note that the survey data collection for this study was conducted in 2021, prior to the Nigerian government’s decision to remove fuel subsidies \citep{adetayo_nigeria_2023}. This policy change has led to a substantial increase in diesel and petrol prices, which is expected to raise the operational costs of diesel generators. Consequently, this price hike may diminish the appeal of diesel generators among businesses and households and increase the attractiveness of electricity from renewable sources \citep{evans2023socio}.

\subsection{Heterogeneous preference for the management and maintenance model}

We found the preference for the model of electricity service management and maintenance to be heterogeneous among both households and SMEs. This lack of a distinct preference among households can be attributed to historical inconsistencies in power supply. Negative experiences with both state-owned electricity companies and private entities, including bureaucratic hurdles and high bills, may have contributed to households' uncertainty about different management models \citep{otobo_critical_nigeria_elecpolicy_2023, alkhuzam_worldbank_private_vs_state_2018}. Additionally, the idea of community-run electricity services is unfamiliar to many households, and issues such as lack of social cohesion exist \citep{ogunleye2022stakeholder}. Thus, such challenges could explain the lack of expressed preference for the community-run electricity model.

On the other hand, businesses are primarily focused on ensuring stable and reliable electricity sources to support their operations. The financial impact of power outages often pushes businesses towards self-sufficient energy solutions like diesel generators or solar systems \citep{leahy2019economic}. However, the own generation of power is costly. Therefore there is a sufficient willingness to pay for formal power if supplies could be assured \citep{roy_breaking_cycle_2023}. These contrasting forces may result in a lack of uniform inclination toward any specific electricity service management model among SMEs.

\subsection{Socioeconomic factors influencing households' willingness to pay}

In Table \ref{tab:interaction} we explored the influence of eight different socioeconomic factors on the household's willingness to pay for electricity connection fees. Taking the factor "weekly expenditure per member" as a proxy for household income, the lower aversion of households with higher expenditure can be associated with their higher purchasing power relative to their lower-income counterparts. This is also observed in other WTP studies, for instance in Tanzania \citep{wen_off-grid_2023} and Ethiopia \citep{aweke_valuing_2022}.

We observe a decline in WTP among older household heads, inline with previous WTP studies in Nigeria \citep{nduka2021get,nduka2023reducing}. The logic may be similar to the weekly expenditure explained above. According to the Nigerian National Bureau of Statistics \citeyearpar{nigerian_national_stats_2011} data, households in which the adults are in their prime are more likely to earn more money from a wider range of income sources than others in the same context. This implies that older households heads, who have passed their prime, earn less and are less resilient to economic shocks (since their income sources are less diversified) than younger households. Consequently, they have a lower WTP.

\begin{figure}
    \centering
    \includegraphics[width=0.7\linewidth]{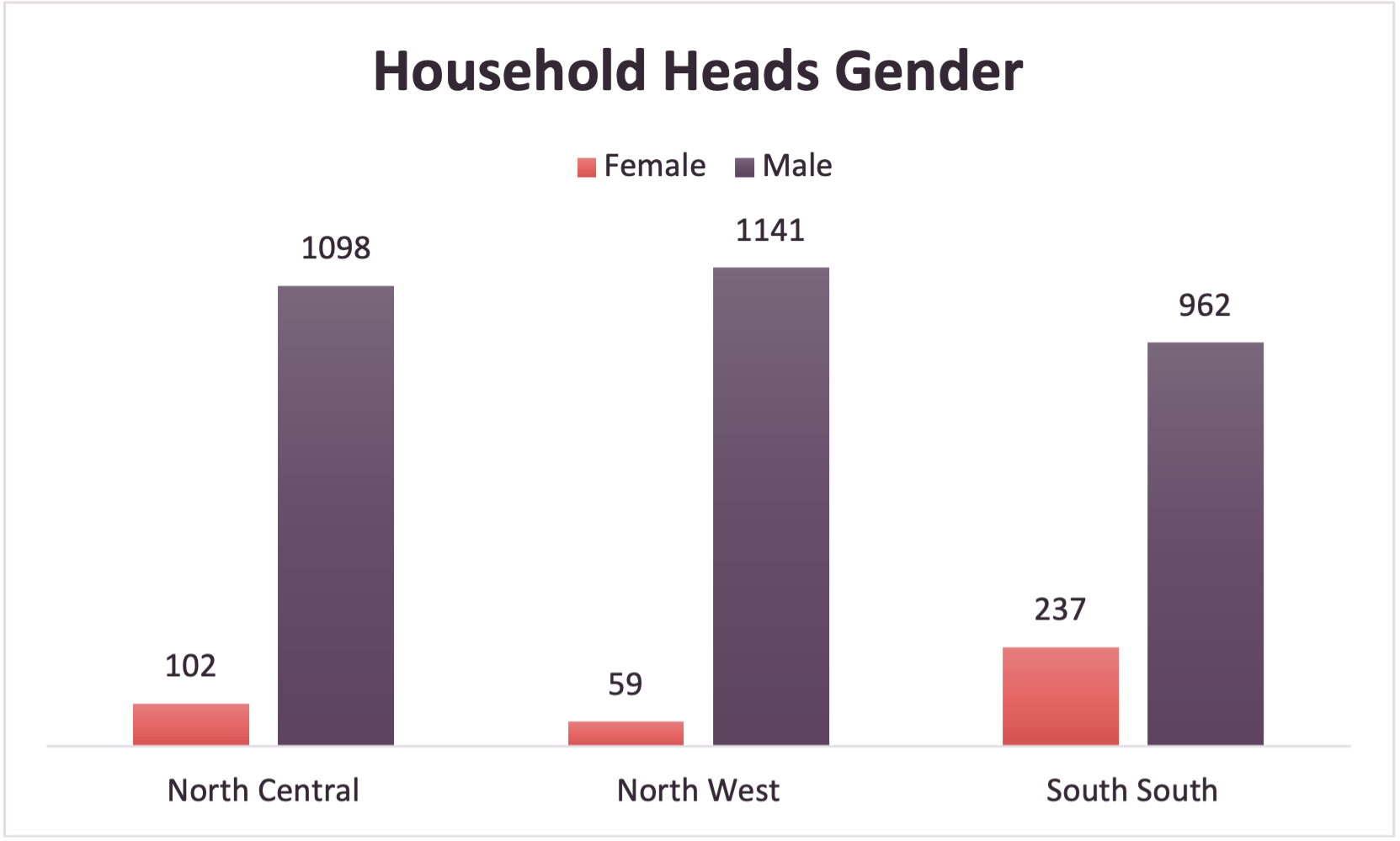}
    \caption{Number of surveyed households with female and male heads in each zone. A total of eleven percent of our households are female-headed across the three zones.}
    \label{fig:male_female_percentage}
\end{figure}

The higher connection fee aversion among households in which the woman has to obtain permission before buying clothes may also have an economic dimension to it. As shown by \cite{sesan2012navigating} in an ethnographic study of household decision-making around cooking energy in rural Kenya, seemingly basic necessities, including food, may be out of reach for the poorest households, who would most benefit from energy access interventions. \\

\subsection{Gender of household heads}

In this study, we are limited to households with either male or female heads. Female-headed households constitute a small share of all the households in our sample (11\% across all three zones; 9\% in the North Central, 5\% in the North West, and 20\% in the South South; see Figure \ref{fig:male_female_percentage} for the exact numbers). Although, these figures are reflective of the lower share of female-headed households in the general population: according to the 2018/2019 Nigerian Living Standards Survey, 18.8\% of all households in the country are headed by women, with variations across the zones \citep{nigerian_national_stats_2020}, our data might not be representative of all female-headed households in rural and peri-urban Nigeria. Nevertheless, our findings can help bring to light some less-investigated aspects of gender in willingness to pay. 

Our finding regarding the higher preference of male-headed households for increased nighttime electricity can be in line with the established trends showing gendered differences in household energy use patterns \citep{grunewald2020societal,shrestha2020gender,shrestha2021review}. Previous studies (e.g., \citealp{wong2009climate}) have found that men in certain contexts may value nighttime electricity for the potential it has to increase their opportunities for leisure after dark - for example, by enabling them to watch television. Women, on the other hand, may simply use the extended hours of lighting to cram more chores into their day, perhaps making them less inclined to prioritize nighttime electricity relative to their male counterparts. 

The absence of a significant gender disparity in our results beyond an increased male preference for nighttime electricity indicates that, in Nigerian rural and peri-urban contexts at least, not having men - who are traditionally assumed to be the primary income earners in those contexts \citep{akanle2020changing} - as household heads may not significantly decrease overall demand for electricity within the population. This is a promising finding for electricity service providers, whether public or private, as it suggests the presence of somewhat uniform latent demand for the service even in contexts where consumers are typically less affluent. On a less positive note, the disparity in spending between male- and female-headed households (as shown in Table \ref{tab:expenditure}) underscores the ongoing presence of the gender pay gap. This gap closely mirrors the national average of 73 percent among individuals engaged in paid employment \citep{nwaka2016gender}. It is therefore important to pay attention to broader social and economic policies that will elevate the earning power of women, which will, in turn, boost their expenditure on a range of household needs, including electricity services.

In discussing our observations, we have to note that the female-headed households in our sample have different characteristics in each zone and that the share of female-headed households is not the same in the three zones (see Figure \ref{fig:male_female_percentage}). We should consider the heterogeneity of female-headed households in our sample when stratified into zones. More insight on gender aspects of PeopleSuN project data in combination with qualitative interviews is presented in \cite{Lich2023}.

\section{Conclusion and policy recommendations} \label{sec:conclusion}
This research focused on the analysis of the results from a discrete choice experiment carried out in a survey of 3,599 households and 1112 SMEs in Nigeria, out of urban cores, with access to grid, and across three distinct geopolitical zones: North West, North Central, and South South. We established several conditional logit models for households and SMEs and calculated their marginal willingness to pay (mWTP) for different electricity service attributes in the upfront connection fee. These mWTP values could inform the pricing strategies of electricity providers venturing into rural Nigerian regions. Furthermore, our models allowed for an investigation into the preferences of households and SMEs concerning electricity service attributes, as well as the effect of socioeconomic factors on these preferences.

In our analysis of preferences for daytime and nighttime electricity, we found that households place a higher value on nighttime electricity, whereas enterprises prioritize electricity during daytime hours. This divergence provides an opportunity for electrification projects to use flexible demand schemes \citep{ruokamo2019towards} and prevent surplus supply. By strategically redistributing the demand peak of enterprises to coincide with the dip in household demand during the day, and conversely, aligning the enterprise demand dip with the household demand peak at night, electrification initiatives can optimize resource allocation and operational effectiveness.

We observed a notable increase in the mWTP of SMEs to obtain high-capacity electricity instead of medium- and low-capacity, thus enabling the operation of a broader spectrum of appliances for them. Conversely, households exhibited a comparatively modest increase in willingness to pay for the same transition. This difference signifies a strategic insight: the development of cost-effective mid-capacity electricity solutions for rural electrification may find greater acceptance among households compared to SMEs. Furthermore, we noted no distinct SME preference for the source of electricity. On the other hand, households exhibited a strong aversion to diesel generators and an appeal toward electricity from the grid or solar solutions. However, the preference for solar solutions instead of grid electricity was not significant for the households. This finding supports rural electrification strategies suggesting to use a mix of electricity solutions to speed up rural electrification in Nigeria \citep{martinot2000regulatory}.

Regarding the influence of socioeconomic factors, we found households with higher expenditure per member and younger household heads to have a higher willingness to pay for electricity connection. Although we did not observe a significant influence of household head gender on their mWTP, male-headed households had a higher preference for nighttime electricity compared to female-headed households. Furthermore, households in which the lead women of the household have to ask for permission before buying clothes for themselves showcased a stronger aversion to elevated electricity fees. These findings are instrumental in tailoring rural electrification policies to bridge social disparities effectively. 

Finally, it is noteworthy that the initial cost of connection remains a barrier for poorer households in Nigeria despite significant capital subsidies that have been made available to offset this cost, most notably in the World Bank-financed Nigeria Electrification Project (NEP) \citep{world_bank_nigeria_electricity_2023}. This indicates that there may be a need for multiple levels of intervention to raise the affordability threshold for those households - not just on the supply side, as is currently being done, but also on the demand side, through a range of innovative payment and consumer financing models.

\section*{Acknowledgements}

This research is a component of the three-year research and development initiative “People Power: Optimizing off-grid electricity supply systems in Nigeria” (PeopleSuN). PeopleSuN is a project supported by the German Federal Ministry of Education and Research (BMBF) under the funding program “Client II - International Partnerships for Sustainable Innovations” (FKZ 03SF0606A). 

We would like to thank PeopleSuN project leads, Clara Neyrand and Philipp Blechinger for their kind support. The present and former colleagues in the PeopleSuN team at RLI, Gregory Ireland, Tobias Rieper, Ulli Lich, Hedwig Bartels, Pierre-Francois Duc, Katrin Lammers, Saeed Sayadi, Setu Pelz and others who played a significant role in this project. Special thanks to our PeopleSuN partners, Ifeoma Malo, Ruth Atsegboa-Mohammed, Abel Gaiya, Maria Yetano, Sophia Schneider, Hannes Cramer, Tapan Kumar, Diego Garcia and Peter Adelmann who attended 2023 Munich Intersolar conference and PeopleSuN retreat in Munich, their feedback during discussion sessions helped us determine the best focus points for our further analysis.

PJ gratefully acknowledges the generous funding from the European Union through the Erasmus Mundus scholarship. He also appreciates the kind efforts of the "MSc. Transition, Innovation, and Sustainability Environments (TISE)" program team during the period he worked on parts of this project for his master's thesis.

\section*{CRediT authorship contribution statement}
\textbf{Pouya Janghorban}: Conceptualization of this study,
Methodology, Formal analysis, Software, Discussions, Writ-
ing - Original Draft, Writing - Review \& Editing. \textbf{Temilade
Sesan}: Discussions, Writing - Review \& Editing. \textbf{Muhammad-
Kabir Salihu}: Discussions, Writing - Review \& Editing.
\textbf{Olayinka S. Ohunakin}: Discussions, Writing - Review
\& Editing. \textbf{Narges Chinichian}: Conceptualization of this
study, Supervision, Discussions, Writing - Original Draft,
Visualization, Writing - Review \& Editing.

\singlespacing
\setlength\bibsep{0pt}
\bibliographystyle{apacite}
\bibliography{references.bib}

\end{document}